\documentclass[10pt, conference]{IEEEtran}

\usepackage{amsmath,amssymb,amsfonts}
\usepackage{algorithmic}
\usepackage{cite}
\usepackage{graphicx}
\usepackage{textcomp}
\usepackage{subcaption}
\usepackage[svgnames]{xcolor}
\usepackage{multirow}
\usepackage[hidelinks]{hyperref}
\usepackage{makecell}
\usepackage{booktabs}
\usepackage{listings}
\usepackage{tikz}
\usepackage{pifont}
\usepackage{comment}
\usepackage{enumitem}
\usepackage[most]{tcolorbox}
\usepackage{scalerel}
\usepackage{makecell}
\usepackage{caption}
\usepackage{float}

\usepackage{longtable}
\usepackage{arydshln}

\usepackage[all=normal,paragraphs=tight,floats=tight,mathspacing=tight,wordspacing=tight,leading=tight,mathdisplays=tight]{savetrees}
\usepackage[stretch=0,shrink=40]{microtype}
\usepackage{setspace}
\definecolor{codegreen}{rgb}{0,0.6,0}
\definecolor{codegray}{rgb}{0.5,0.5,0.5}
\definecolor{codepurple}{rgb}{0.58,0,0.82}
\definecolor{codebgcolor}{rgb}{0.95,0.95,0.92}
\lstdefinestyle{codestyle}{
    backgroundcolor=\color{codebgcolor},
    commentstyle=\color{codegreen},
    keywordstyle=\color{magenta},
    numberstyle=\tiny\color{codegray},
    stringstyle=\color{codepurple},
    basicstyle=\linespread{0.8}\ttfamily\footnotesize,
    breakatwhitespace=false,         
    breaklines=true,                 
    captionpos=b,                    
    keepspaces=true,                 
    showspaces=false,                
    showstringspaces=false,
    showtabs=false,                  
    tabsize=2,
    escapechar=\%,
    columns=fullflexible,
    framexleftmargin=4pt,
}
\lstdefinestyle{promptstyle}{
    basicstyle=\linespread{0.8}\ttfamily\footnotesize,
    breakatwhitespace=true,         
    breaklines=true,
    captionpos=b,                    
    keepspaces=true,                 
    showspaces=false,                
    showstringspaces=false,
    showtabs=false,                  
    tabsize=2,
    breakautoindent=True
    escapechar=\%,
    columns=fullflexible,
}

\lstnewenvironment{python}{\lstset{style=codestyle,language={Python}}}{}
\lstnewenvironment{prompt}{\lstset{style=promptstyle,language={}}}{}

\newcommand*\emptycirc[1][1ex]{\tikz\draw (0,0) circle (#1);} 
\newcommand*\halfcirc[1][1ex]{%
  \begin{tikzpicture}
  \draw[fill] (0,0)-- (90:#1) arc (90:270:#1) -- cycle ;
  \draw (0,0) circle (#1);
  \end{tikzpicture}}
\newcommand*\fullcirc[1][1ex]{\tikz\fill (0,0) circle (#1);}

\tcbset{colback=blue!1!white,colframe=blue!60!black,fontupper=\small,breakable,size=small}

\newcommand{\ourwork}{{HLS-Eval}}

\title{\huge\ourwork: A Benchmark and Framework for Evaluating LLMs on High-Level Synthesis Design Tasks}

\author{
Stefan Abi-Karam\textsuperscript{1,2}, Cong Hao\textsuperscript{1}
\\
Georgia Institute of Technology\textsuperscript{1}, Georgia Tech Research Institute\textsuperscript{2}
\\
\{\href{mailto:stefanabikaram@gatech.edu}{\nolinkurl{stefanabikaram}}, \href{mailto:callie.hao@gatech.edu}{\nolinkurl{callie.hao}}\}@gatech.edu}

\begin{document}
\bstctlcite{IEEEexample:BSTcontrol}

\maketitle

\begin{abstract}
The rapid scaling of large language model (LLM) training and inference has accelerated their adoption in semiconductor design across academia and industry. Most prior works benchmark LLMs for design tasks involving hardware description languages (HDLs), primarily Verilog. Meanwhile, designers are increasingly using high-level synthesis (HLS) to develop domain-specific accelerators and other hardware systems. However, benchmarks and tooling to comprehensively evaluate LLMs for HLS design tasks remain scarce.

To address this, we introduce HLS-Eval, the first comprehensive benchmark and evaluation framework for LLM-driven HLS design tasks. HLS-Eval focuses on evaluating two key high-level tasks: 1) generating HLS code from natural language descriptions, and 2) making HLS-specific code edits to existing HLS code, primarily targeting HW optimization and performance gains. To evaluate these tasks, we construct the HLS-Eval benchmark consisting of 94 unique designs as benchmark cases. These designs are drawn from a diverse mix of established community HLS benchmarks and novel sources. Using a semi-automated flow, we prepare each case to be "LLM-ready", supplemented with a natural language description and a corresponding testbench for C-simulation and HLS synthesis validation.

Beyond the benchmark designs, HLS-Eval provides a framework for automated, parallel evaluation of both local and hosted LLMs across various HLS design tasks. It features a parallel evaluation engine, seamless HLS tool interfaces for LLMs, and an abstraction to support different LLM interaction paradigms, all wrapped in a modular Python API. With HLS-Eval, users can rapidly prototype and evaluate new benchmark sources, HLS design tasks, and LLM methodologies, speeding up the research and development of new AI-driven HLS workflows.

We demonstrate the utility of HLS-Eval through baseline evaluations of popular open-source LLMs for HLS code generation and code editing tasks targeting Vitis HLS. Our evaluation assesses LLM-generated HLS code across four critical metrics: "parseability", "compilability", "runnability", and "synthesizability" — mirroring the iterative debugging and testing process of an HLS designer. Additionally, we provide pass@k metrics for common design tasks, establishing clear evaluation criteria, baselines, and the necessary infrastructure for others in the LLM hardware design community to build upon.

We open-source our benchmark, framework, and evaluation results at \textcolor{blue}{\url{https://github.com/stefanpie/hls-eval}}.
\end{abstract}

\section{Introduction}

Large language models (LLMs) have demonstrated the ability to model both natural language text and structured computer programs \cite{llm_code}. Recent work has evaluated LLMs' capabilities for hardware design, primarily by using LLMs to write code for hardware description languages (HDLs) \cite{hdleval}, particularly Verilog \cite{verigen, verilogeval, verilogeval_v2, verilog_tool_feedback}, supported by developments in LLM-focused HDL datasets \cite{mg_verilog, rtl_repo}. While significant effort has focused on LLM-based methods for HDL design, little attention has been given to applying LLMs for high-level synthesis (HLS) design.

Unlike traditional HDL-based development, HLS requires distinct domain expertise \cite{hls_expert_design}. Most HLS tools accept a restricted subset of C++ that excludes dynamic memory allocation, complex pointer types, recursion, and standard library features. To enable efficient hardware generation, designers must structure loops, control flow, and other programming constructs to meet compiler requirements (e.g., perfectly nested loops, compile-time loop bounds, static array sizes, single-producer single-consumer patterns). HLS tools also provide specialized hardware directives, known as "pragmas", to guide optimization (e.g., loop unrolling, pipelining, memory partitioning, dataflow streaming). In addition, some HLS tools supply C++ libraries for specialized hardware components, such as arbitrary fixed-point data types, fixed-point math functions, FIFO streams, and task-based dataflows. The specific subset of supported C++, the available pragmas, compiler quirks, and specialized libraries differ among HLS tool vendors (e.g., Vitis HLS v. Intel HLS Compiler v. Catapult HLS), further fragmenting the domain knowledge needed for effective HLS design.

LLMs offer hope to aid designers in tackling the complexity and domain expertise needed to both write efficient HLS designs "from scratch", or more commonly, porting existing CPU-targeted or GPU-targeted algorithms written in C/C++ to HLS for acceleration and deployment on FPGA and ASIC platforms with the promise of increased performance and energy efficacy. These goals represent both code generation and code editing tasks respectively, with the second task being a common appeal of the democratization of hardware acceleration \cite{democratizing_dsc}. In particular, we define "code editing" not just as making C++ code synthesizable but as performing \textit{hardware-optimizing edits}. This is where the true value of HLS lies, benefiting both domain experts and, with the aid of an LLM, less experienced users.

However, adapting LLMs for use in HLS development flows is still challenging and is relatively unexplored. There is a lack of large, diverse, open-source "LLM-for-HLS" benchmarks that provide "LLM-ready" HLS designs for both code generation and code editing evaluations. Secondary to a lack of benchmarks, there is no extensible software framework for researchers and practitioners to create new LLM-ready HLS benchmarks, integrate HLS tool interfaces for LLMs, and run parallel evaluations of new code generation and editing tasks for HLS design. 

In total, tackling both benchmark creation and benchmarking software would democratize the research of "LLMs for HLS" itself. By providing the components (i.e., "LLM-ready" HLS benchmarks, LLM model abstractions, LLM tool interfaces to HLS tools, common HLS task prompts, ...), we enable others in the community, both in academia and industry, to quickly prototype emerging LLM research ideas applicable to HLS design (e.g., domain-specific HLS benchmarks, LLM HLS agents, RAG over HLS knowledge sources, inference-time scaling with HLS tools as verifiers, etc.).

Therefore, we propose \textbf{\ourwork{}} as a comprehensive benchmark and an evaluation framework of LLMs for HLS design tasks. We summarize our contributions as follows:

\setlist[itemize]{leftmargin=1em}
\begin{itemize}

    \item \textbf{Designs Benchmark}: We present a comprehensive evaluation benchmark with 94 HLS designs sourced from community HLS benchmarks, academic textbooks, and open-source hardware accelerators. Each design is "LLM-ready," including a testbench, detailed natural language description, and reference implementation, all of which are manually reviewed.

    \item \textbf{Extensible Evaluation Framework}: We present an open-source Python-based HLS evaluation framework which integrates \textbf{HLS tool interfaces} for LLMs, local and remote \textbf{LLM inference}, and modular \textbf{Evaluator API} abstraction allowing for user-defined inference flows.
    
    \item \textbf{Parallel Evaluation Engine}: To accelerate benchmarking with LLM inference and HLS tool calls, we propose a fine-grained parallel evaluation engine so users can fully exploit their compute resources and speed up evaluations.
    
    \item \textbf{Baseline Task Evaluation}: We evaluated our benchmark with several open-source LLMs to gather initial pass@k metrics for both HLS code generation and HLS code editing tasks. This includes unbiased pass@k for validating whether generated designs can be parsed, can be compiled, can successfully execute the testbench, and can be synthesized by an HLS tool (Vitis HLS in this work).
    
    \item \textbf{Code Generation \& Code Editing Tasks}: In addition to code generation from natural language descriptions, we propose a set of code \textbf{editing tasks aimed at optimizing hardware design}, going beyond simply non-synthesizable to synthesizable code transformations.
\end{itemize}

\section{Prior Work and Challenges}

There is a relatively small body of literature on the topic of "LLMs for HLS" with three primary works focused on benchmarking LLMs for HLS design: C2HLSC~\cite{c2hlsc_mlcad}, HLSPilot~\cite{hlspilot}, and Gai et al.~\cite{auto_llm_hls}. We directly compare \ourwork{} to these studies in Table \ref{tab:compare-with-existing} to highlight our contributions over existing works.

C2HLSC is an early benchmarking effort motivated by the use case of translating naive C++ to HLS C++. This work has a small, handcrafted benchmark of designs based on cryptographic primitives and functions from the NIST randomness test suite. The authors conduct both manual and automated evaluations of the naïve-to-HLS code conversion task.

HLSPilot extends the concept of transforming naive C++ into optimized HLS C++ one step further. Rather than solely focusing on generating valid, synthesizable C++ code, it introduces inference-time techniques such as retrieval-augmented generation (RAG), chain-of-thought prompting, and automated design space exploration (DSE). There is a stronger focus to generate highly optimized HLS hardware designs using an LLM rather than merely achieving synthesizability. HLSPilot's benchmark  contains designs from the Rosetta~\cite{rosetta} benchmark, along with a few hand-crafted application-focused designs.

Gai et al. evaluates the ability of LLMs to generate kernel implementations from natural language descriptions. They also explore fine-tuning LLMs for this code generation task. This benchmark designs in this work are sourced from HLSyn~\cite{hlsyn} and HLSDataset\cite{hlsdataset}, which are supersets of the PolyBench~\cite{polybench}, MachSuite~\cite{machsuite}, and CHStone~\cite{chstone} benchmarks.

While these works explore different subsets of design tasks and benchmark sources, they remain relatively limited in benchmarking diversity and lack a holistic framework for evaluation. In some cases, such as C2HLSC, the benchmark designs are open-source and well-structured, but the overall benchmark set is small and not diverse. On the other hand, Gai et al. include a larger pool of benchmark designs with natural language descriptions for each design, but their work is currently not open-source. Moreover, these existing studies are not designed to be holistic evaluation frameworks but rather target specific designer use cases, such as maximizing hardware optimization in HLSPilot.

Overall, there remains a need for a holistic and modular approach to HLS-based LLM evaluation—one that is extensible by researchers and practitioners to contribute new benchmark designs, design tasks, and inference techniques. Additionally, there is a need to aggregate and curate the existing corpus of high-quality published / open-source HLS designs into a benchmark set that includes the appropriate metadata and code structure (e.g., descriptions and functional testbenches) for effective LLM evaluation. We hope that \ourwork{} can address these needs as an open-source benchmark and evaluation framework for the broader research community.

\begin{table}[t]
    \centering
        \caption{A comparison of \ourwork{} with the existing work. \fullcirc{}: feature supported; \emptycirc{}: feature unsupported; \halfcirc{}: feature partially supported; "Opt.": Optimization}
    \setlength\tabcolsep{2pt}
    \renewcommand{\arraystretch}{0.8}
    \resizebox{\columnwidth}{!}{%
    \begin{tabular}{l|c|c|c|c}
        \toprule
        \textbf{Contributions}         & \textbf{C2HLSC} & \textbf{HLSPilot} & \textbf{Gai et.\ al} &
        
        \textbf{\underline{\ourwork}} \\
        \midrule
        Benchmark --- Polybench             & \emptycirc                            & \emptycirc                            &  \fullcirc  & \fullcirc               \\
        Benchmark --- MachSuite             & \emptycirc                            & \emptycirc                           &  \fullcirc   & \fullcirc               \\
        Benchmark --- CHStone               & \emptycirc                            & \emptycirc                         &    \fullcirc    & \fullcirc               \\
        Benchmark --- Rosetta               & \emptycirc                            & \fullcirc                         &    \emptycirc    & \fullcirc               \\
        Academic --- PP4FPGA              & \emptycirc                            & \emptycirc                          &   \emptycirc     & \fullcirc               \\
        Academic --- C2HLSC              & \fullcirc                            & \emptycirc                          &   \emptycirc     & \fullcirc               \\
        Open-Source --- Accelerators & \emptycirc                    & \emptycirc                              &  \emptycirc  & \fullcirc               \\
        \hdashline
        Total Benchmark Cases & 10                    & 19                              &  52  & 94               \\
        \hline
        Tool --- C-Simulation              & \fullcirc & \fullcirc                  &      \fullcirc        & \fullcirc \\
        Tool --- HLS Synthesis                  & \fullcirc & \fullcirc                     &     \fullcirc     & \fullcirc \\
        \hline
        Tasks --- Code Generation                  & \emptycirc & \emptycirc                     &     \fullcirc     & \fullcirc \\
        Tasks --- Opt. Driven Code Editing                   & \halfcirc & \fullcirc                     &     \emptycirc     & \fullcirc \\
        \hline
        Local LLM Inference  & \emptycirc & \emptycirc                     &     \fullcirc     & \fullcirc \\
        Remote LLM Inference    & \fullcirc & \fullcirc                     &     \emptycirc     & \fullcirc \\
        \hline
        Parallel Benchmarking Engine              & \emptycirc & \emptycirc                  &      \emptycirc        & \fullcirc \\
        \hline
        Extensible Python API & \emptycirc & \emptycirc                    &     \emptycirc       & \fullcirc \\
        Open-Source & \fullcirc & \fullcirc                    &     \emptycirc       & \fullcirc \\
        \bottomrule
    \end{tabular}%
    }

    \label{tab:compare-with-existing}
\end{table}

\section{\ourwork{} Framework}

\begin{figure}[t!]
    \centering
    \includegraphics[width=0.4\textwidth]{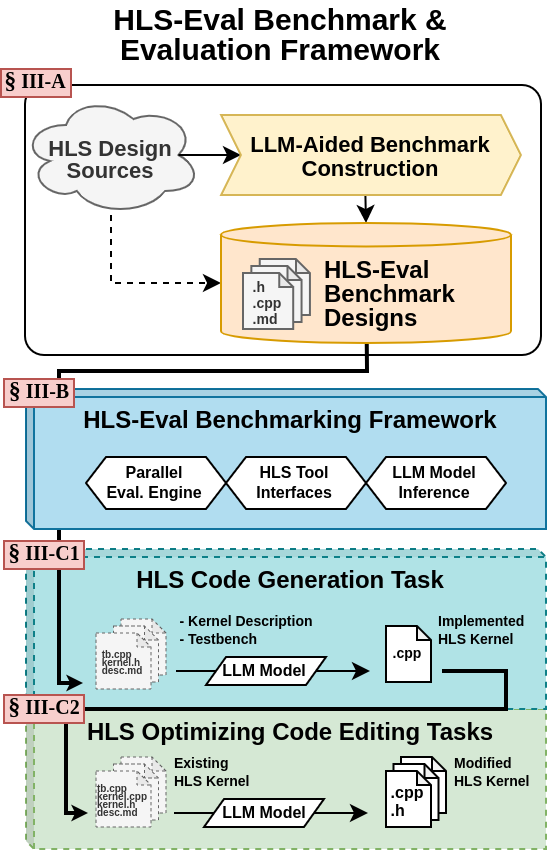}
    \caption{Overview of \ourwork{}, which includes benchmark construction, an evaluation framework, and HLS design tasks such as code generation and optimization-based code editing evaluations.}
    \label{fig:main_overview}
\end{figure}

As shown in Figure \ref{fig:main_overview}, we organize \ourwork{} into three parts: the \ourwork{} benchmark construction (\S\ref{sec:benchmark_designs}), the \ourwork{} software framework for parallel end-to-end modular evaluations (\S\ref{sec:eval_framework}), and evaluations for both HLS code generation (\S\ref{sec:eval_code_gen}) and hardware-optimizing HLS code editing tasks (\S\ref{sec:eval_code_edit}). 

\subsection{\textbf{Benchmark Construction}}
\label{sec:benchmark_designs}

\begin{table}[t!]
    \centering
    \resizebox{\columnwidth}{!}{
        \begin{tabular}{l|ccccc}
            \toprule
            \textbf{\makecell{Data\\Source}} & \textbf{\makecell{\# of Bench.\\Designs}} & \textbf{\makecell{Average\\Kernel LoC}} & \textbf{\makecell{Average HLS\\ Synthesis Runtime}} \\
            \midrule
            \textbf{Polybench \cite{polybench}} & $28$ & $\approx23$ & $\approx43$ s. \\
            \textbf{MachSuite \cite{machsuite}} & $17$ & $\approx59$ & $\approx54$ s. \\
            \textbf{CHStone \cite{chstone}} & $20$ & $\approx70$  & $\approx49$ s.   \\
            \textbf{Rosetta \cite{rosetta}} & $8$ & $\approx17$  & $\approx44$ s.   \\
            \textbf{C2HLSC \cite{c2hlsc_mlcad}} & $12$   & $\approx47$  & $\approx38$ s.   \\
            \textbf{PP4FPGA \cite{pp4fpga}} & $3$ & $\approx36$  & $\approx39$ s.   \\
            \textbf{FlowGNN \cite{flowgnn_conf}} & $3$  & $\approx36$  & $\approx48$ s.   \\
            \textbf{GNNBuilder \cite{gnnb_conf}} & $3$   & $\approx23$  & $\approx38$ s.   \\
            \midrule
            \textbf{Totals} & $94$ Designs & $4048$ LoC & $\approx38$ m.   \\
            \bottomrule
        \end{tabular}
    }
    \caption{Description of "LLM-ready" designs in the \ourwork{} benchmark. "LoC": lines-of-code, "Average HLS Synthesis Runtime": evaluated with Vitis HLS 2024.1 as a proxy for design complexity and average delay of HLS tool in an LLM evaluation loop.}
    \label{tab:benchmark_data}
\end{table}

To evaluate HLS design tasks, we construct a benchmark set of 94 HLS designs as summarized in Table \ref{tab:benchmark_data}.
We primarily source HLS designs from an existing work, HLSFactory \cite{hlsfactory_conf}. HLSFactory includes designs from common HLS community benchmarks, primarily PolyBench~\cite{polybench}, CHStone~\cite{chstone}, MachSuite~\cite{machsuite}, and Rosetta~\cite{rosetta}, each targeting a range of applications, including scientific computing / HPC kernels, digital signal processing, cryptography, floating-point computations, and deep learning acceleration. Additionally, we include designs from other sources, such as the academic textbook \textit{Parallel Programming for FPGAs}~\cite{pp4fpga}, as well as open-source neural network accelerators, including FlowGNN~\cite{flowgnn_conf} and GNNBuilder~\cite{gnnb_conf}. Finally, we incorporate the benchmark cases from C2HLSC~\cite{c2hlsc_mlcad}, mostly related to cryptography and NIST randomness tests.

During the construction of the \ourwork{} benchmark, we found that most HLS source code lacks standardized organization and sufficient metadata to be useful for LLM research, i.e., “LLM-ready.” For example, PolyBench kernels contain many unexpanded C++ macros and unnecessary utility code. MachSuite kernels are written with HLS in mind but lack critical metadata, such as a detailed natural language description of the kernel, and require additional harness code for testbenches that are unnecessary for synthesis. Some designs, such as those from Rosetta or larger deep learning accelerators, are overly complex, with code spread across multiple source files and headers, making them impractical as individual LLM benchmark cases for this work (though they remain valuable to evaluate LLMs for hierarchical HLS design tasks in the future).

Ideally, our goal is to create a benchmark with "LLM-ready" designs, where each design has the following elements:

\setlist[itemize]{leftmargin=1em}
\begin{itemize}
    \item A single header file containing typedefs, define statements, macros, constant data/arrays, and the kernel function signature of the top-level HLS kernel.
    \item A single C++ file that implements the kernel, including a top function and, if necessary, any sub-functions and specialized data types.
    \item A natural language description of the kernel, detailing the inputs, outputs, and relevant specifications of the kernel's design or operation.
    \item A self-contained C++ testbench with testcases / checking to compare against the kernel execution.
\end{itemize}

\begin{figure}[t!]
    \centering
    \includegraphics[width=0.45\textwidth]{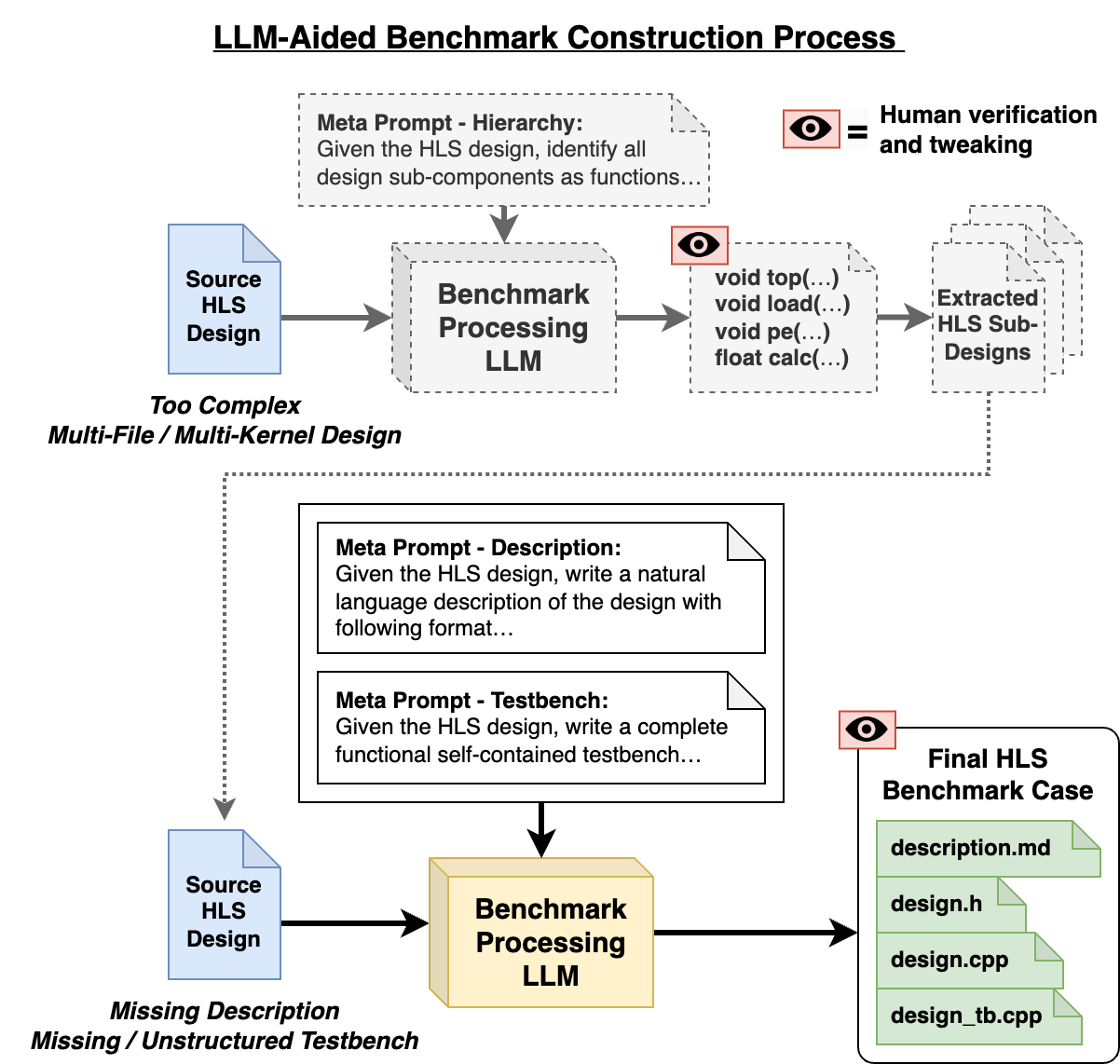}
    \caption{Overview of the LLM-aided benchmark construction given some arbitrary unstructured HLS source code. Note the human-in-the-loop for manually tweaking and review of the final benchmark design.}
    \label{fig:bench-collection}
\end{figure}

To achieve the goal of translating designs from various sources into these “LLM-ready” designs, we built a semi-automated, human-in-the-loop workflow that utilizes LLMs to aid in the benchmark construction process. We refer to this meta-workflow as “LLM-aided benchmark construction,” which is illustrated in Figure \ref{fig:bench-collection}. In this workflow, we leverage open-source LLMs along with predefined meta-prompts to assist researchers in transforming arbitrarily formatted, often unstructured, HLS source code into “LLM-ready” benchmark cases that are subsequently added to \ourwork{}.

As such, we developed and used this workflow as a CLI tool integrated into \ourwork{} with three main commands:

\setlist[itemize]{leftmargin=1em}
\begin{itemize}
    \item \textbf{Hierarchy Extraction}: Given arbitrary HLS source code and a defined top function, identify all other sub-functions and hardware components in the C++ source code and extract them into separate C++ files.
    \item \textbf{Description Generation}: Given HLS source code and a defined top function, generate a structured natural language description of the kernel, detailing its high-level functionality and any specific algorithmic details relevant to the implementation. Additionally, include structured metadata such as the function signature, data types, constants, and subcomponents in the design.
    \item \textbf{Testbench Generation}: Given HLS source code, a defined top function, and an optional description, generate a C++ testbench that verifies the functionality of the HLS kernel, returning 0 if correct and 1 if any test cases fail.
\end{itemize}

Use of these tools is optional, as not all features are required for every case. For example, PolyBench and MachSuite designs already include testbenches and primarily require manual tweaking and simplification of scaffolding code, with minimal use of the LLM-aided description generator. However, CHStone required full use of hierarchy extraction, description generation, and testbench generation to translate CHStone designs into benchmark cases. Additionally, Hierarchy Extraction enables users to create multiple HLS benchmark cases from a single large design, contributing to a more diverse benchmark set.

Using this CLI tool, we curated the final \ourwork{} benchmark set of designs, manually verifying and refining the benchmarks along the way. We utilized a combination of \texttt{Llama 3.3 70B Instruct}, \texttt{Llama 3 70B Instruct}, and \texttt{Qwen2.5 Coder 32B Instruct} models. The final set of designs is packaged with \ourwork{}’s code, allowing end-users to load the built-in \ourwork{} benchmark designs as well as integrate their own benchmark designs at runtime.

\subsection{\textbf{Evaluation Framework}}
\label{sec:eval_framework}

\subsubsection{\textbf{HLS Tool Interfaces for LLMs}}
\label{sec:tools}

To evaluate LLM-generated code and enable LLM-HLS tool interaction, we provide an “LLM-friendly” Python API for AMD/Xilinx’s Vitis HLS. This abstraction supports two HLS tool interfaces: \texttt{VitisHLSCSimTool} and \texttt{VitisHLSSynthTool}. Fig.~\ref{listing:tools} in Section \ref{sec:example_tools} of the Appendix shows an example standalone usage of these tool interfaces.

Each tool interface processes source files, auxiliary files, and flow-specific arguments to execute either C-simulation or HLS synthesis. \texttt{VitisHLSCSimTool} follows a two-step process: first, it compiles LLM-generated HLS code and testbenches (\texttt{csim\_design -setup}) without execution; if successful, it runs the compiled test bench binary (\texttt{csim.exe}). This allows for differentiation of syntax / compilation errors from functional correctness errors. \texttt{VitisHLSSynthTool} runs standard HLS synthesis (\texttt{csynth\_design}), lowering C++ designs to RTL.
Each tool invocation captures return codes, standard output, and errors. This execution metadata provides targeted feedback for iterative LLM refinement, supporting complex workflows and LLM-driven design iterations.

\subsubsection{\textbf{LLM Model Interface}}
\label{sec:model_interface}

To support both local and remote inference for various LLMs, we integrate the \texttt{vLLM} and \texttt{simonw/llm} Python libraries. \texttt{vLLM} enables local inference, leveraging available GPU or accelerator resources and supports open-source LLMs from HuggingFace. \texttt{simonw/llm} facilitates remote inference on hosted models, including commercial (e.g., OpenAI, Anthropic) and open-source (e.g., Together AI) offerings. For evaluations, we primarily use Together AI’s hosted open-source models. These frameworks are encapsulated within our \texttt{Model} abstraction.

\subsubsection{\textbf{Parallel and Customizable HLS Evaluation}}
\label{sec:eval_backend_and_api}

\begin{figure}[t!]
    \centering
    \includegraphics[width=0.45\textwidth]{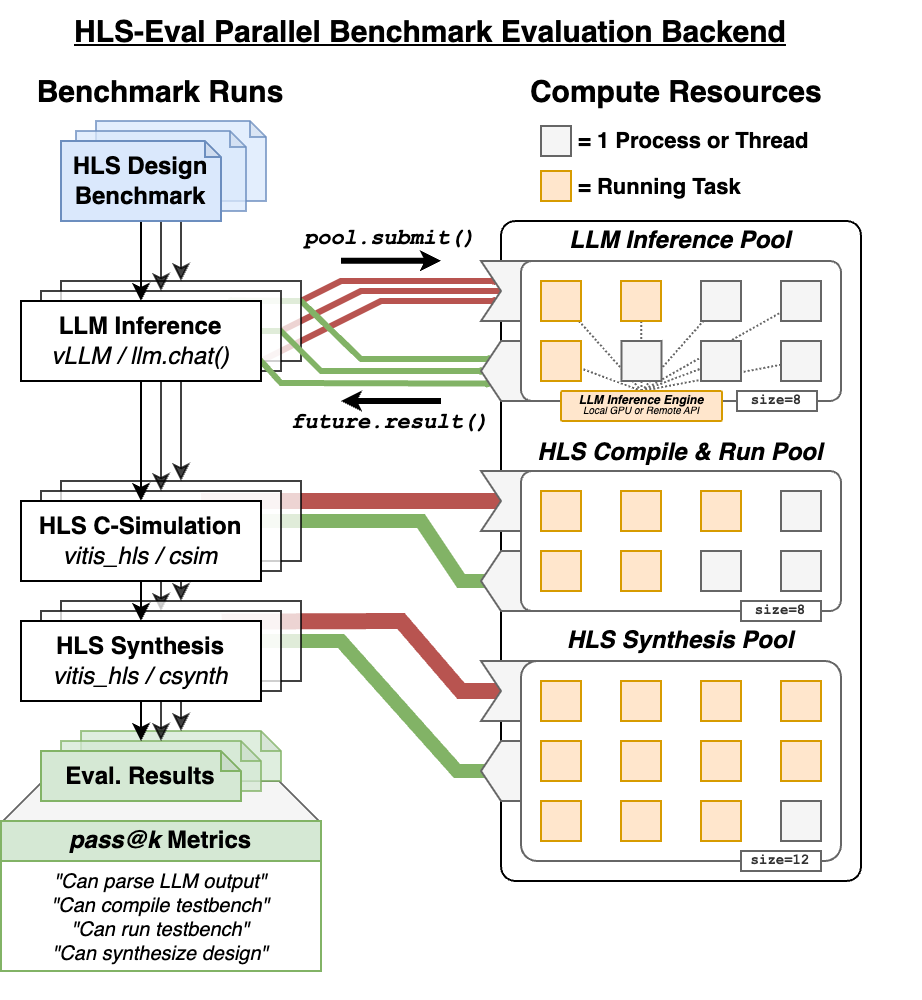}
    \caption{The parallel benchmark evaluation backend for \ourwork{}. Execution at each stage of the deigns flow can be submitted to a fixed-sized pool of threads on a server or user workstation. Different stages can have different sized pools to set different parallelisms for each stage to maximize utilization of compute resources}
    \label{fig:eval-engine}
\end{figure}

Scaling LLM evaluations across multiple benchmark cases requires efficient parallel execution. For example, in a zero-shot HLS code generation workflow, each benchmark involves LLM inference, C-simulation, and HLS synthesis, each taking seconds to minutes per design and model. Efficient parallelization is essential for practical scalability, especially when incorporating tool-feedback iterations or inference-time techniques like RAG, hierarchical prompting, and custom reasoning.

To enable flexible and scalable evaluations, \ourwork{} introduce a "Parallel Evaluation Engine" and an \texttt{Evaluator} API. The evaluation engine parallelizes LLM inference and HLS tool execution across job pools, while the \texttt{Evaluator} abstraction allows defining custom inference flows independent of specific designs or models. Together, these components enable fast, extensible evaluations of HLS design tasks. Figure \ref{listing:simple_eval_example} in Section \ref{sec:example_eval} of the Appendix illustrates a minimal \ourwork{} script for evaluating zero-shot code generation on the full benchmark set.

\subsubsection{\textbf{Parallel Evaluation Engine}}

\begin{figure}[t!]
    \centering
    \includegraphics[width=0.45\textwidth]{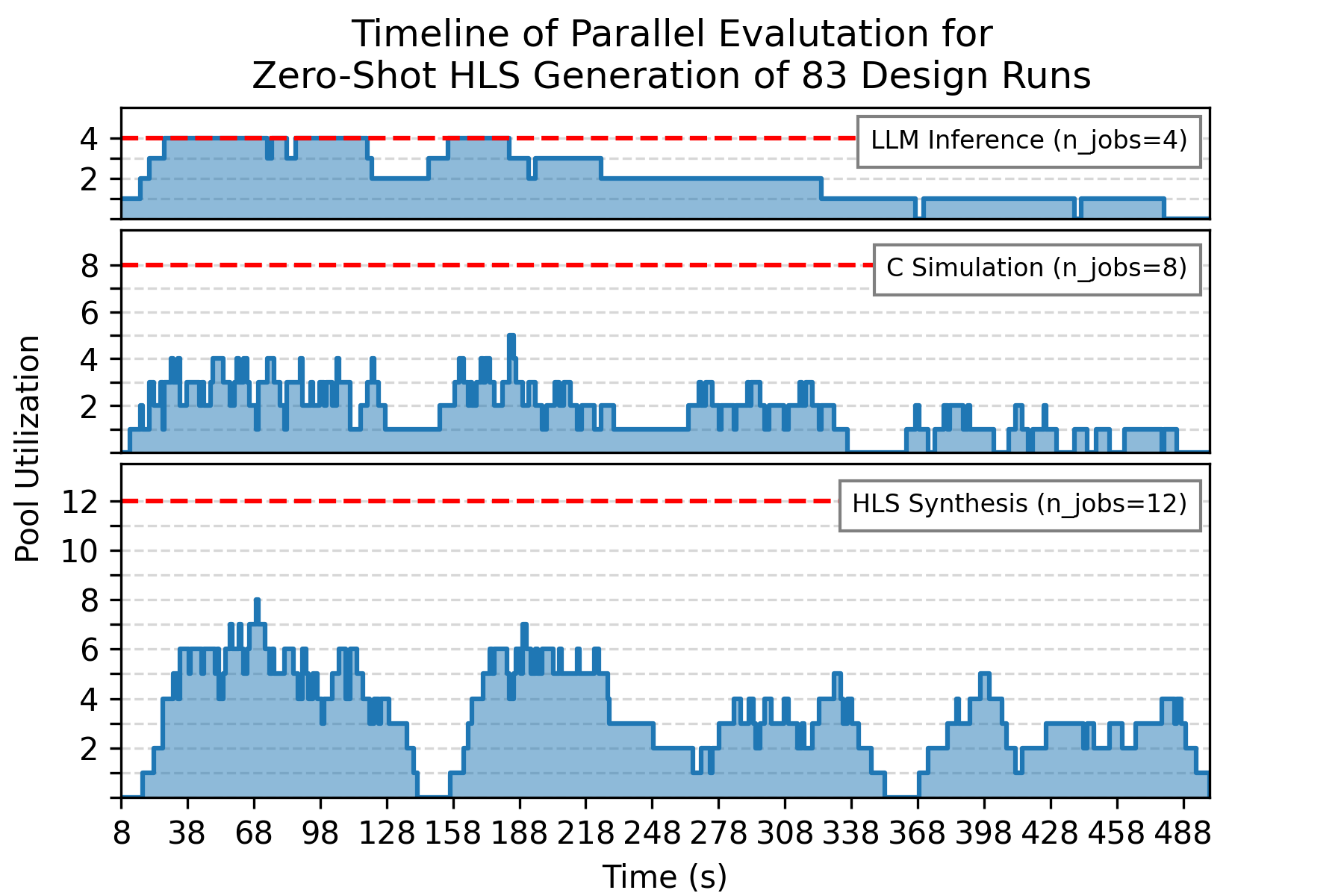}
    \caption{An execution trace of the \ourwork{} parallel benchmark elevation backend. The graphs show the pool utilization of all tasks pools over time. The red line shows the user set pool size for each task.}
    \label{fig:eval-timeline}
\end{figure}

As shown in Figure~\ref{fig:eval-engine}, the parallel evaluation engine executes HLS design tasks with fine-grained user-defined parallelism for LLM inference and HLS tools. In a zero-shot generation workflow, LLM inference would run first, followed by HLS C-simulation, and finally HLS synthesis, repeated for each benchmark design. To accelerate evaluations, multiple designs are processed in parallel, enabling simultaneous inference calls and HLS executions.

A naive approach assigns each benchmark design to a separate evaluation thread, executing LLM inference and HLS tools sequentially. However, this method is bottlenecked by HLS synthesis, which has significantly longer runtimes than LLM inference or C-simulation. With multiple threads, inference completes quickly, but all threads may stall on synthesis, leaving resources underutilized.

To mitigate this, \ourwork{} employs fine-grained parallelization. Users define the number of parallel evaluation threads while independently tuning parallelism for LLM inference, HLS C-simulation, and synthesis. Each thread submits tasks to dedicated thread pools, aka "task pools". These pools (implemented via queues) process tasks in the order they were submitted, ensuring efficient execution without bottlenecks. Figure~\ref{fig:eval-timeline} illustrates task pool utilization during execution.

This approach provides four tunable parallelism factors: \texttt{n\_jobs} (parallel evaluation threads), \texttt{n\_jobs\_llm}, \texttt{n\_jobs\_csim}, and \texttt{n\_jobs\_synth}. Users can balance workloads across their hardware limitations, optimizing resource utilization across all evaluation stages. 

\subsubsection{\textbf{Evaluator API}}

To support diverse design tasks and model-agnostic LLM inference, \ourwork{} defines an \texttt{Evaluator} API. Each \texttt{Evaluator} requires users to implement an \texttt{evaluate\_design(benchmark\_case, model)} function, handling LLM invocation, code extraction, HLS tool execution, and metrics logging. This function implementation is the only requirement for integrating a new evaluator type. For parallel evaluations across multiple designs and models, all \texttt{Evaluator} objects implement \texttt{evaluate\_designs(benchmark\_cases, models, ...)}, which automatically applies \texttt{evaluate\_design} across all benchmark-model combinations. Each \texttt{Evaluator} is initialized with an \texttt{EvalThreadPools} object, managing the parallel evaluation engine, along with \texttt{VitisHLSCSimTool} and \texttt{VitisHLSSynthTool} for simulation and synthesis.

Currently, we provide two benchmark evaluators: \textit{HLSGenerationZeroShotEvaluator} and \textit{HLSEditingZeroShotEvaluator}, supporting zero-shot code generation and editing. We plan to release tool-feedback evaluators in the near-future. Users can also define custom evaluators for new HLS design tasks (e.g., testing, verification) and inference techniques (e.g., RAG, hierarchical design) as they see fit.

\subsection{\textbf{Evaluation Design Tasks}}
\label{sec:evals}

\begin{figure}
\centering
\vspace{-24pt}
\begin{subfigure}[t]{0.45\textwidth}
    \centering
    \includegraphics[width=\linewidth]{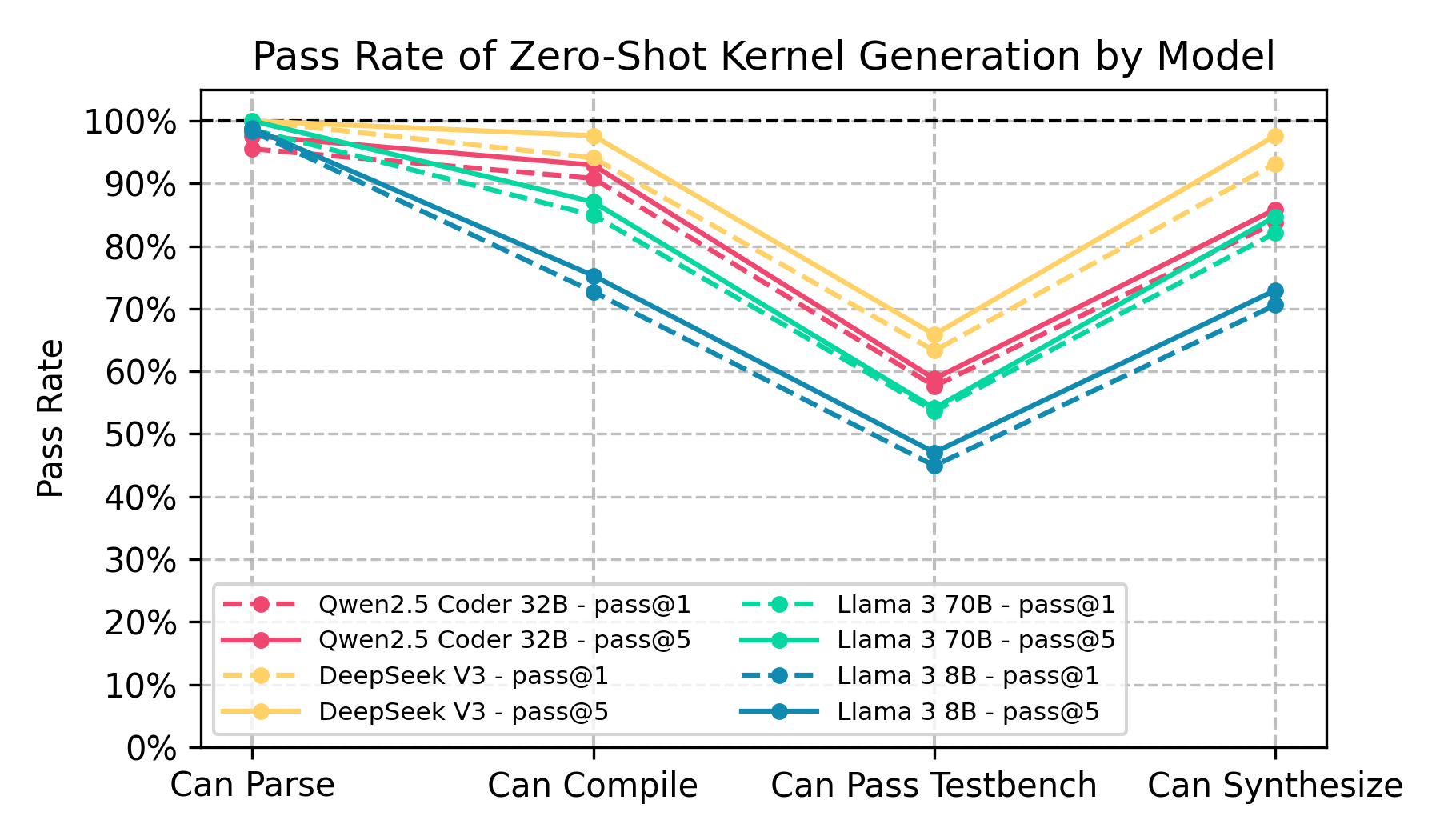} 
    \vspace{-48pt}
    \label{fig:res_gen}
\end{subfigure}

\begin{subfigure}[t]{0.45\textwidth}
    \centering
    \includegraphics[width=\linewidth]{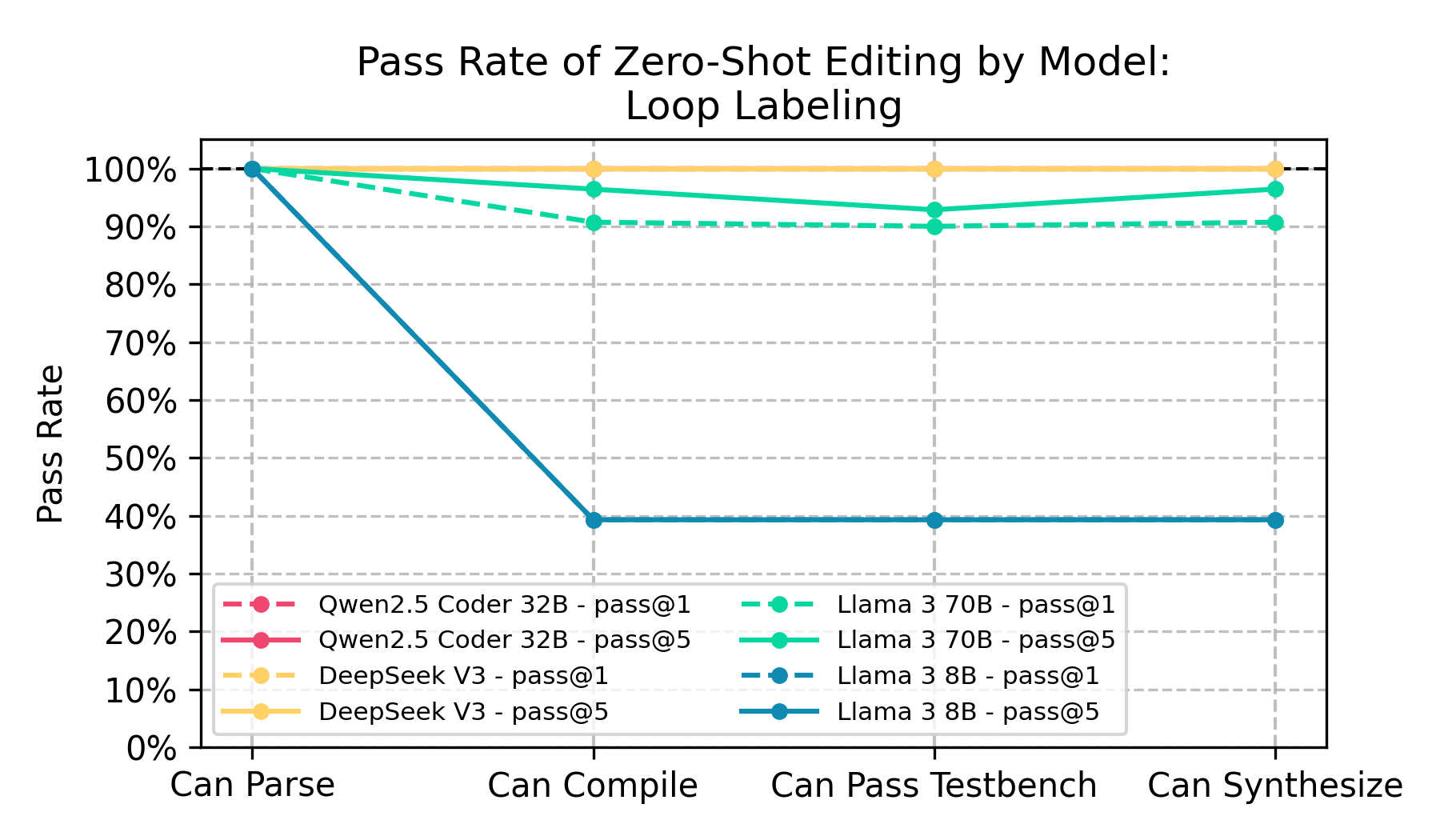}
    \vspace{-48pt}
    \label{fig:res_label}
\end{subfigure}

\begin{subfigure}[t]{0.45\textwidth}
    \centering
    \includegraphics[width=\linewidth]{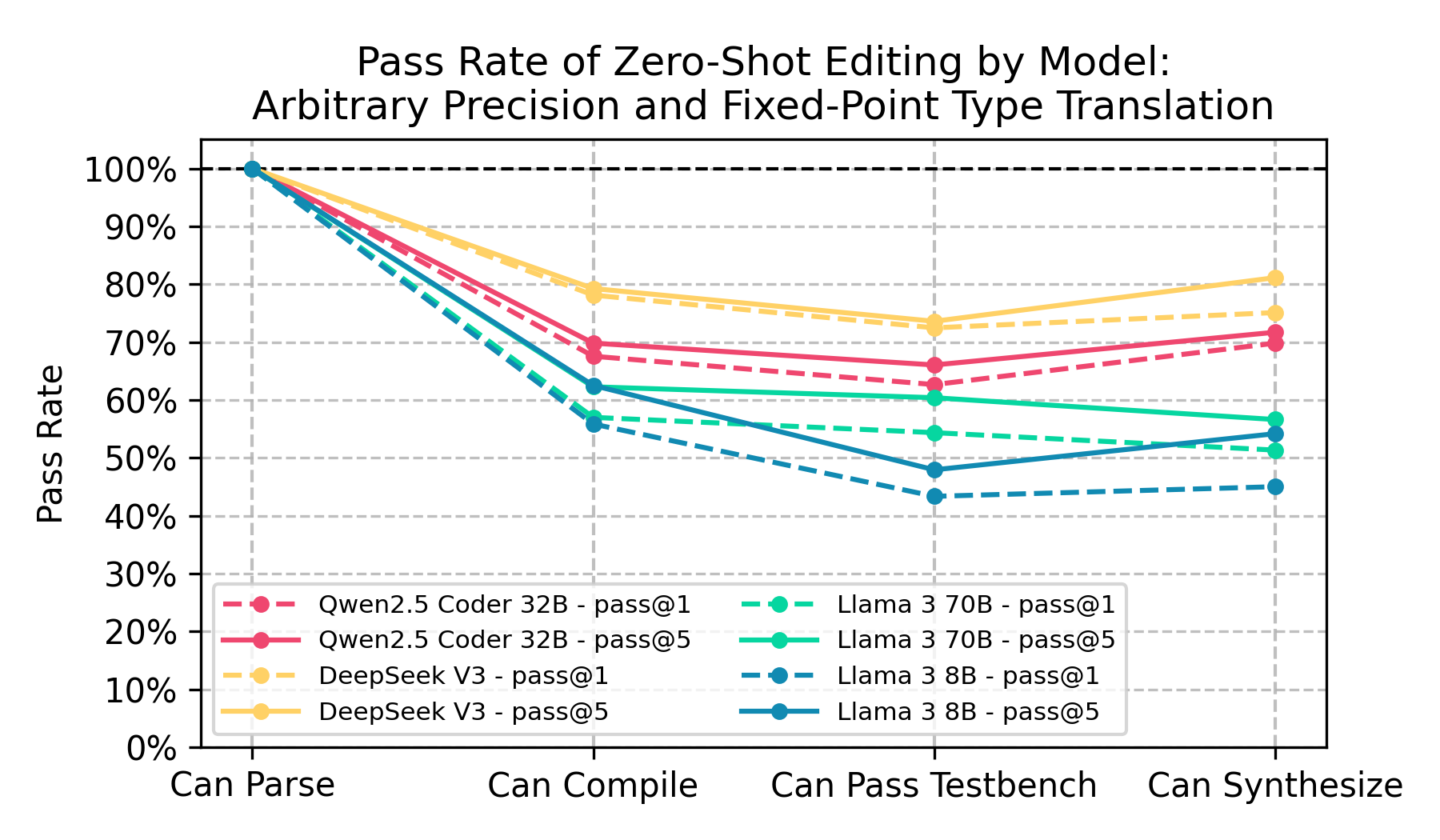} 
    \vspace{-48pt}
    \label{fig:res_fpx}
\end{subfigure}

\begin{subfigure}[t]{0.45\textwidth}
    \centering
    \includegraphics[width=\linewidth]{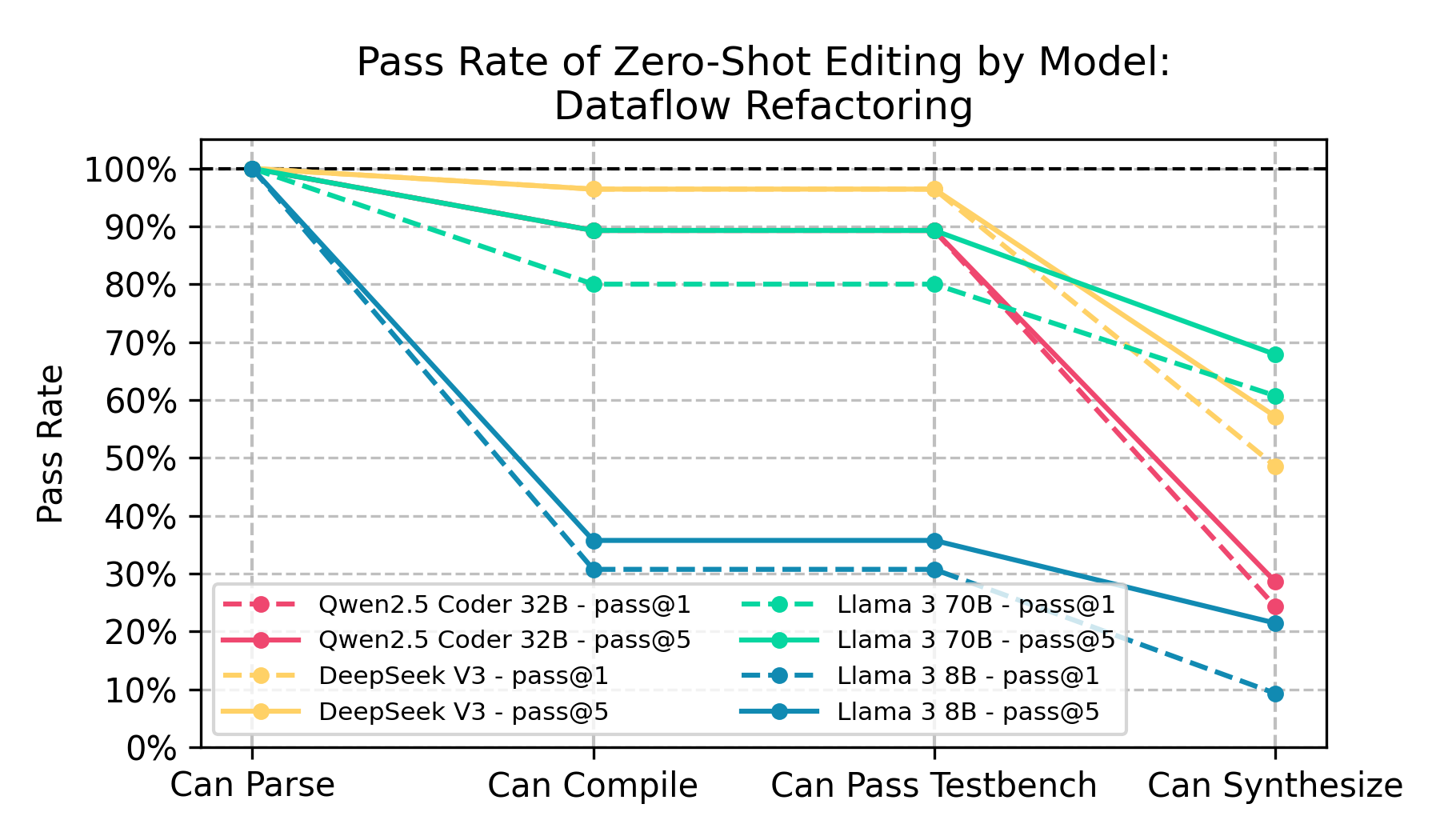}
    \vspace{-48pt}
    \label{fig:res_dataflow}
\end{subfigure}

\begin{subfigure}[t]{0.45\textwidth}
    \centering
    \includegraphics[width=\linewidth]{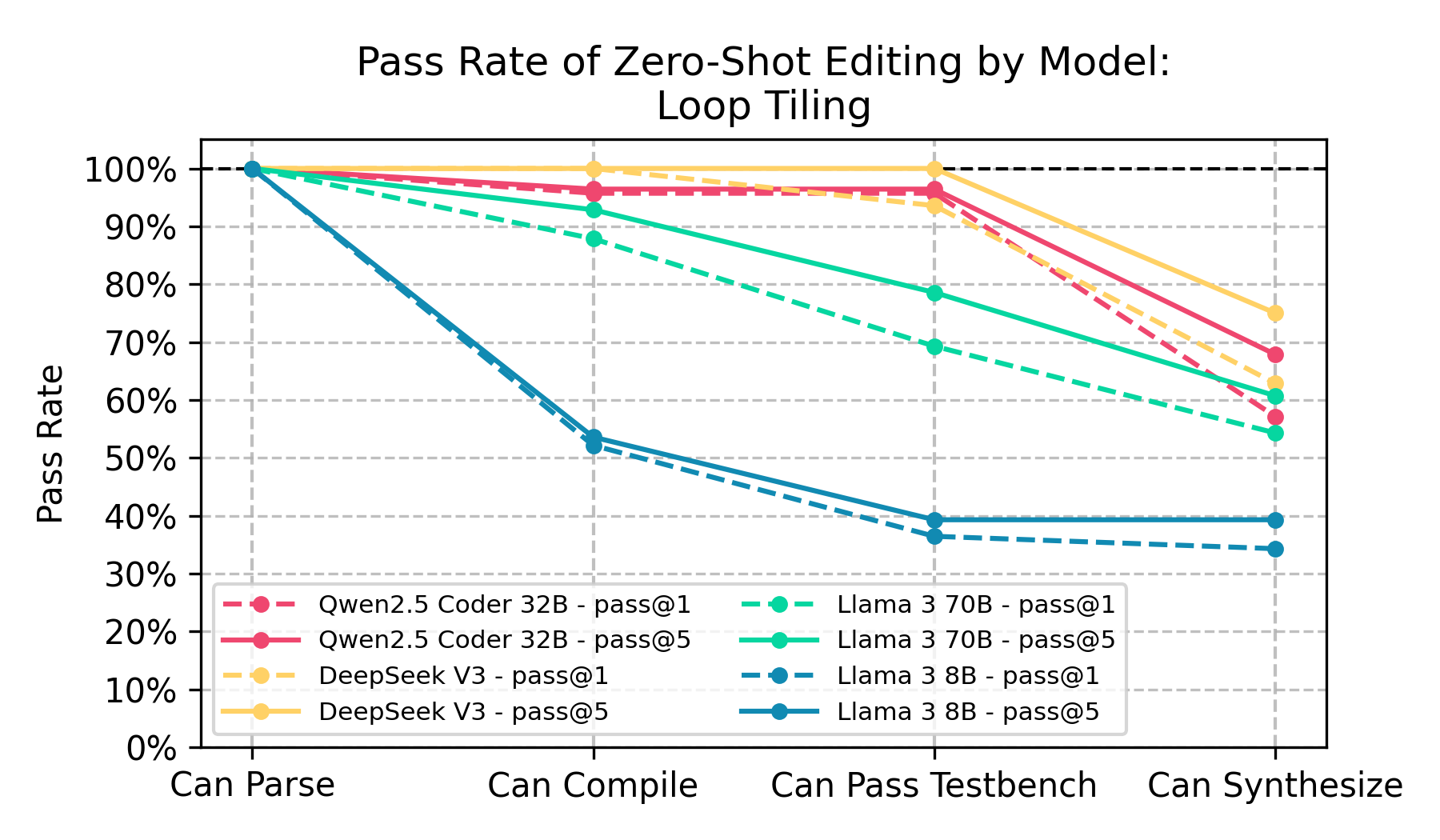} 
    \vspace{-48pt}
    \label{fig:res_tile}
\end{subfigure}

\caption{Evaluation results of HLS code generation and HLS code editing tasks; "Pass Rate": average pass@k over all evaluated designs for a given design stage and model}
\label{fig:mega-fig}

\end{figure}

To demonstrate the utility of \ourwork{} and establish a reference baseline performance applying open-source models to HLS design tasks, we present evaluation results for both code generation and various code editing tasks.

For each task, we assess LLM-generated or LLM-edited HLS code using the following metrics:

\begin{itemize}
    \item \textbf{Parseability} – The ability to extract code blocks from the LLM output in the expected format.
    \item \textbf{Compilability} – The ability of the HLS tool’s C++ compiler to successfully compile the generated code and testbench.
    \item \textbf{Runnability} – The ability of the compiled HLS code and testbench to execute and pass all test cases (i.e., return an exit code of 0).
    \item \textbf{Synthesizability} – The ability of the generated HLS code to be synthesized by the HLS tool.
\end{itemize}

For each benchmark design, these metrics are evaluated as either pass or fail; this data is aggregated to compute an unbiased pass@k value~\cite{llm_code} for each HLS-specific metric.

All evaluations are conducted using AMD/Xilinx Vitis HLS 2024.1 for C-simulation and HLS synthesis. We evaluate four open-source instruction-tuned models: \texttt{Llama 3 70B}, \texttt{Llama 8B}, \texttt{Qwen 2.5 Coder 32B}, and \texttt{DeepSeek V3}. For all evaluations, we use a temperature parameter of $T=0.7$ and sample $N=5$ responses per design, unless otherwise specified. In this manner, we can compute unbiased pass@k metrics for $k=5$ and $k=1$.

All results in this section are shown in Figure \ref{fig:mega-fig} and presented as tables in Section \ref{sec:tables} of the Appendix.

\subsubsection{\textbf{HLS Code Generation}}

\label{sec:eval_code_gen}

The most straightforward code generation task for HLS design is generating the HLS kernel implementation from a natural language description. For this task, we provide the model with a prompt containing a natural language description of the kernel, the header file for the kernel, and the testbench code used for evaluation. We consider this a realistic evaluation scenario, as HLS designers often start with an initial higher-level model of their algorithm they wish to implement, allowing them to generate a testbench with known test cases for evaluation.

We evaluate this task on the subset of PolyBench, MachSuite, CHStone, Rosetta, and C2HLSC designs.

\subsubsection{\textbf{Optimization driven HLS Editing}}
\label{sec:eval_code_edit}
We present several HLS code editing tasks in our baseline evaluation. With the exception of loop labeling as an "easy baseline" task, the remaining tasks focus on code edits that attempt to \textit{optimize} the hardware implementation: arbitrary precision and fixed-point type translation, data flow refactoring, and loop tiling.

\textbf{Adding Loop Labels}: Adding loop labels to existing HLS code improves readability and makes it easier to track how the HLS tool processes loop regions. For example, labeled loops appear clearly in reports with user-defined identifiers, as opposed to being assigned automatically generated identifiers which are harder to interpret. Loop labels also facilitate defining pragmas that target loops and make it easier to specify a design space with unrolling and pipelining directives using user-defined labels.

This is one of the simplest editing tasks, as it does not require functional or non-local modifications. We simply prompt the LLM to rewrite the code by adding loop labels to all loops in the design.

We evaluate this task on the subset of PolyBench designs where this optimization is most applicable.

\textbf{Arbitrary Precision and Fixed-Point Type Translation}: When accelerating computation-heavy kernels using floating-point types or bit manipulation, designers may optimize the design (e.g. lower latency and area) by switching to arbitrary precision (AP) integer and fixed-point types. While this task may seem like a simple "search-and-replace" refactoring of the code, AP integer and fixed-point translation is non-trivial, requiring understanding the existing types in the design, determining appropriate sizing and precision, and potentially utilizing vendor-specific fixed-point libraries for both types and mathematical operations.

In this task, the goal is to refactor the design by replacing AP integer and floating-point types with fixed-point types from Vitis HLS, specifically \texttt{ap\_int<...>}, \texttt{ap\_uint<...>}, \texttt{ap\_fixed<...>}, and modifying mathematical operations to use Vitis HLS AP integer fixed-point functions when necessary (e.g., \texttt{hls::exp(...)}).

We evaluate this task on the subset of PolyBench, CHStone, and C2HLSC designs where this optimization is most applicable.

\textbf{Dataflow Refactoring}: HLS tools can perform dataflow optimizations when subfunctions and the data passed between them adhere to specific structural constraints imposed by the HLS tool. Properly designed HLS kernels with dataflow optimizations enable efficient streaming-like data movement and simplify data dispatch to processing elements (PEs).

However, unoptimized kernels may not be inherently "dataflow-friendly" and require refactoring. This refactoring must also comply with compiler-imposed dataflow rules, such as the "single-producer, single-consumer" constraint for variable access. Additionally, designing dataflow-style kernels introduces further complexity, as it requires understanding runtime behavior to model the latency input data-dependent control flow. For this task, we ask the LLM to refactor a given HLS kernel such that the dataflow pragma can be applied effectivly.

We evaluate this task on the subset of PolyBench designs where this optimization is most applicable.

\textbf{Loop Tiling}: Loop tiling and unrolling are common HLS optimizations that improve parallelism and memory access efficiency, particularly in loop-heavy code such as linear algebra and stencil-based kernels found in scientific computing. Loop tiling partitions a loop’s iteration space into smaller blocks, enhancing data locality and reducing memory bandwidth bottlenecks. Loop unrolling replicates loop iterations, minimizing loop control overhead and increasing instruction-level parallelism.

For this task, we prompt the LLM to refactor the given HLS code by applying manual loop tiling source code transformations and unrolling directives where appropriate. The modified code should use pragmas such as \textit{\#pragma HLS UNROLL} for unrolling and \textit{\#pragma HLS ARRAY\_PARTITION} to align with tiling when applicable, optimizing memory accesses. The refactored design should ensure that tiling and unrolling do not introduce loop dependencies that would cause initiation interval (II) violations.

We evaluate this task on the subset of PolyBench designs where this optimization is most applicable.

\section{Conclusion}

\ourwork{} introduces a much-needed benchmark and software framework for researchers and designers exploring the application of LLMs for HLS design. Our baseline evaluations demonstrate the effectiveness of \ourwork{} in achieving this goal on a smaller scale while also providing the community with a larger, more diverse set of "LLM-ready" HLS designs for further evaluation and exploration of LLM inference techniques.

Looking ahead, we plan to extend \ourwork{} by incorporating a broader range of benchmark designs and integrating tool-feedback evaluators.

By open-sourcing \ourwork{}, we aim to foster collaboration within the research hardware design community, inviting contributions and growing the field of LLM-aided HLS design. 

\bibliographystyle{IEEEtran}
\bibliography{refs}

\newpage
\onecolumn

\appendix

\subsection{Evaluation Results Tables}
\label{sec:tables}

\begin{table*}[h!]
    \centering
    \setlength{\tabcolsep}{2pt}

\begin{tabular}{l|rr|rr|rr|rr}
\cmidrule[\heavyrulewidth]{2-9}
 & \multicolumn{2}{c|}{\textbf{Can Parse}} & \multicolumn{2}{c|}{\textbf{Can Compile}} & \multicolumn{2}{c|}{\textbf{Can Pass TB}} & \multicolumn{2}{c}{\textbf{Can Synth}} \\
\midrule
 \textbf{Model} & \textbf{pass@1} & \textbf{pass@5} & \textbf{pass@1} & \textbf{pass@5} & \textbf{pass@1} & \textbf{pass@5} & \textbf{pass@1} & \textbf{pass@5} \\
\midrule
DeepSeek V3 & 100.0\% & 100.0\% & 94.1\% & 97.6\% & 63.3\% & 65.9\% & 93.2\% & 97.6\% \\
Qwen2.5 Coder 32B & 95.5\% & 97.6\% & 90.8\% & 92.9\% & 57.6\% & 58.8\% & 83.8\% & 85.9\% \\
Llama 3 70B & 98.6\% & 100.0\% & 84.9\% & 87.1\% & 53.6\% & 54.1\% & 82.1\% & 84.7\% \\
Llama 3 8B & 98.4\% & 98.8\% & 72.7\% & 75.3\% & 44.9\% & 47.1\% & 70.6\% & 72.9\% \\
\bottomrule
\end{tabular}

    \caption{Zero-Shot Eval. Results for HLS Kernel Generation Task}
    \label{tab:tab_gen}
\end{table*}

\begin{table*}[h!]
    \centering
    \setlength{\tabcolsep}{2pt}
\begin{tabular}{l|rr|rr|rr|rr}
\cmidrule[\heavyrulewidth]{2-9}
 & \multicolumn{2}{c|}{\textbf{Can Parse}} & \multicolumn{2}{c|}{\textbf{Can Compile}} & \multicolumn{2}{c|}{\textbf{Can Pass TB}} & \multicolumn{2}{c}{\textbf{Can Synth}} \\
\midrule
 \textbf{Model} & \textbf{pass@1} & \textbf{pass@5} & \textbf{pass@1} & \textbf{pass@5} & \textbf{pass@1} & \textbf{pass@5} & \textbf{pass@1} & \textbf{pass@5} \\
\midrule
DeepSeek V3 & 100.0\% & 100.0\% & 100.0\% & 100.0\% & 100.0\% & 100.0\% & 100.0\% & 100.0\% \\
Qwen2.5 Coder 32B & 100.0\% & 100.0\% & 100.0\% & 100.0\% & 100.0\% & 100.0\% & 100.0\% & 100.0\% \\
Llama 3 70B & 100.0\% & 100.0\% & 90.7\% & 96.4\% & 90.0\% & 92.9\% & 90.7\% & 96.4\% \\
Llama 3 8B & 100.0\% & 100.0\% & 39.3\% & 39.3\% & 39.3\% & 39.3\% & 39.3\% & 39.3\% \\
\bottomrule
\end{tabular}
    \caption{Zero-Shot Eval. Results for HLS Editing Task --- Loop Labeling}
    \label{tab:tab_label}
\end{table*}

\begin{table*}[h!]
    \centering
    \setlength{\tabcolsep}{2pt}

\begin{tabular}{l|rr|rr|rr|rr}
\cmidrule[\heavyrulewidth]{2-9}
 & \multicolumn{2}{c|}{\textbf{Can Parse}} & \multicolumn{2}{c|}{\textbf{Can Compile}} & \multicolumn{2}{c|}{\textbf{Can Pass TB}} & \multicolumn{2}{c}{\textbf{Can Synth}} \\
\midrule
 \textbf{Model} & \textbf{pass@1} & \textbf{pass@5} & \textbf{pass@1} & \textbf{pass@5} & \textbf{pass@1} & \textbf{pass@5} & \textbf{pass@1} & \textbf{pass@5} \\
\midrule
DeepSeek V3 & 100.0\% & 100.0\% & 78.1\% & 79.2\% & 72.5\% & 73.6\% & 75.1\% & 81.1\% \\
Qwen2.5 Coder 32B & 100.0\% & 100.0\% & 67.5\% & 69.8\% & 62.6\% & 66.0\% & 69.8\% & 71.7\% \\
Llama 3 70B & 100.0\% & 100.0\% & 57.0\% & 62.3\% & 54.3\% & 60.4\% & 51.3\% & 56.6\% \\
Llama 3 8B & 100.0\% & 100.0\% & 55.8\% & 62.5\% & 43.3\% & 47.9\% & 45.0\% & 54.2\% \\
\bottomrule
\end{tabular}

    \caption{Zero-Shot Eval. Results for HLS Editing Task --- Arbitrary Precision and Fixed-Point Type
Translation}
    \label{tab:tab_fpx}
\end{table*}

\begin{table*}[h!]
    \centering
    \setlength{\tabcolsep}{2pt}

\begin{tabular}{l|rr|rr|rr|rr}
\cmidrule[\heavyrulewidth]{2-9}
 & \multicolumn{2}{c|}{\textbf{Can Parse}} & \multicolumn{2}{c|}{\textbf{Can Compile}} & \multicolumn{2}{c|}{\textbf{Can Pass TB}} & \multicolumn{2}{c}{\textbf{Can Synth}} \\
\midrule
 \textbf{Model} & \textbf{pass@1} & \textbf{pass@5} & \textbf{pass@1} & \textbf{pass@5} & \textbf{pass@1} & \textbf{pass@5} & \textbf{pass@1} & \textbf{pass@5} \\
\midrule
DeepSeek V3 & 100.0\% & 100.0\% & 96.4\% & 96.4\% & 96.4\% & 96.4\% & 48.6\% & 57.1\% \\
Qwen2.5 Coder 32B & 100.0\% & 100.0\% & 89.3\% & 89.3\% & 89.3\% & 89.3\% & 24.3\% & 28.6\% \\
Llama 3 70B & 100.0\% & 100.0\% & 80.0\% & 89.3\% & 80.0\% & 89.3\% & 60.7\% & 67.9\% \\
Llama 3 8B & 100.0\% & 100.0\% & 30.7\% & 35.7\% & 30.7\% & 35.7\% & 9.3\% & 21.4\% \\
\bottomrule
\end{tabular}

    \caption{Zero-Shot Eval. Results for HLS Editing Task --- Dataflow Refactoring}
    \label{tab:tab_dataflow}
\end{table*}

\begin{table*}[h!]
    \centering
    \setlength{\tabcolsep}{2pt}

\begin{tabular}{l|rr|rr|rr|rr}
\cmidrule[\heavyrulewidth]{2-9}
 & \multicolumn{2}{c|}{\textbf{Can Parse}} & \multicolumn{2}{c|}{\textbf{Can Compile}} & \multicolumn{2}{c|}{\textbf{Can Pass TB}} & \multicolumn{2}{c}{\textbf{Can Synth}} \\
\midrule
 \textbf{Model} & \textbf{pass@1} & \textbf{pass@5} & \textbf{pass@1} & \textbf{pass@5} & \textbf{pass@1} & \textbf{pass@5} & \textbf{pass@1} & \textbf{pass@5} \\
\midrule
DeepSeek V3 & 100.0\% & 100.0\% & 100.0\% & 100.0\% & 93.6\% & 100.0\% & 62.9\% & 75.0\% \\
Qwen2.5 Coder 32B & 100.0\% & 100.0\% & 95.7\% & 96.4\% & 95.7\% & 96.4\% & 57.1\% & 67.9\% \\
Llama 3 70B & 100.0\% & 100.0\% & 87.9\% & 92.9\% & 69.3\% & 78.6\% & 54.3\% & 60.7\% \\
Llama 3 8B & 100.0\% & 100.0\% & 52.1\% & 53.6\% & 36.4\% & 39.3\% & 34.3\% & 39.3\% \\
\bottomrule
\end{tabular}

    \caption{Zero-Shot Eval. Results for HLS Editing Task --- Loop Tiling}
    \label{tab:tab_tile}
\end{table*}

\subsection{Unbiased pass@k Compuation}

The unbiased pass@k metric from \cite{llm_code} is formulated as follows:

\begin{equation}
\text{pass@}k:=\underset{\text { Problems }}{\mathbb{E}}\left[1-\frac{\binom{n-c}{k}}{\binom{n}{k}}\right]
\end{equation}

Where $n$ is the total number of samples per benchmark case, $c$ is the number of correct or passing samples, and $k$ is the pass rate which you want to compute. Normally $n$ is chosen to be $\ge k$ and the larger $n$ is, the more accurate the estimator is. The value in brackets corresponds to the pass rate for a single benchmark case, which can then be average over the entire benchmark set or selected subsets. See \cite{llm_code} for further justification of using the unbiased estimator as opposed to using $1-(1-\hat{p})^k$.

The metric is computed in a numerically stable way as follows:

\begin{python}
def pass_at_k(n, c, k):
    """
    :param n: total number of samples
    :param c: number of correct samples
    :param k: k in pass@k
    """
    if n - c < k: return 1.0
    return 1.0 - np.prod(1.0 - k / np.arange(n - c + 1, n + 1))
\end{python}

\subsection{Example Benchmark Evaluation Code}
\label{sec:example_eval}

\begin{figure}[h!]
    \centering
    \scriptsize
\begin{python}
all_benchmark_case_dirs = find_benchmark_case_dirs(DIR_HLS_EVAL_DESIGNS)
all_benchmark_cases = [
    BenchmarkCase(d, name=d.name) for d in all_benchmark_case_dirs
]

model_to_test = "Qwen/Qwen2.5-Coder-32B-Instruct"
model = build_model_remote_tai(model_to_test, api_key=API_KEY_TOGETHERAI)

vhls = unwrap(auto_find_vitis_hls_dir())

evaluator = HLSGenerationZeroShotEvaluator(
    vitis_hls_tool_csim=VitisHLSCSimTool(vhls),
    vitis_hls_tool_synth=VitisHLSSynthTool(vhls),
    output_data_dir=DIR_CURRENT_OUTPUT_DATA,
)

evaluator.evaluate_designs(
    all_benchmark_cases,
    [model],
    n_jobs=16,
    n_jobs_pool_llm=4,
    n_jobs_pool_csim=8,
    n_jobs_pool_synth=8,
)
\end{python}
    \caption{The complete code needed using \ourwork{} for simple zero-shot evaluation of HLS generation over all benchmark cases.}
    \label{listing:simple_eval_example}
\end{figure}

\subsection{Example HLS Tool Interface Code}
\label{sec:example_tools}

\begin{figure}[h!]
    \centering
    \scriptsize
\begin{python}
design = Design.from_path("./kenel_gemm/")

csim_tool = VitisHLSCSimTool()
output_compile, output_run = csim_tool.run(
    design.design_dir,
    design.source_files,
    design.not_source_files
)
# data_execution: return code, stdout, stderr
print(output_compile.data_execution)
print(output_run.data_execution)

synth_tool = VitisHLSSynthTool()
synth_output = synth_tool.run(
    design.design_dir,
    design.source_files,
    hls_top_function=design.top_name
)
# data_execution: return code, stdout, stderr
print(synth_output.data_execution)
# HLS Synthesis Report (latency + resources)
print(synth_output.data_tool)
\end{python}
    \caption{Usage of the built-in tool interfaces for Vitis HLS C-Simulation and HLS Synthesis.}
    \label{listing:tools}
\end{figure}

\subsection{Example Design as a \ourwork{} Benchmark Case}
\label{sec:example_design}

The example design below shows the componets of an "LLM-ready" design that is part of the \ourwork{} benchmark. 

\begin{tcolorbox}[title=Example Benchmark Design --- df\_mul64To128]
    \tcbsubtitle{kernel\_description.md}
\begin{prompt}
Kernel Description:
The `mul64To128` kernel is designed to perform a 64-bit by 64-bit multiplication and produce a 128-bit result. The multiplication is broken down into smaller parts to handle the large product size, which is typical in hardware design to manage bit-width limitations and improve efficiency. The algorithm splits each 64-bit input into two 32-bit parts, performs partial multiplications, and then combines these results to form the final 128-bit product. This approach ensures that each multiplication step remains within the 64-bit range, avoiding overflow issues. The kernel handles the carry propagation between the partial products to ensure the correctness of the final result.

---

Top-Level Function: `mul64To128`

Complete Function Signature of the Top-Level Function:
`void mul64To128(bits64 a, bits64 b, bits64 *z0Ptr, bits64 *z1Ptr);`

Inputs:
- `a`: A 64-bit unsigned integer representing the first multiplicand.
- `b`: A 64-bit unsigned integer representing the second multiplicand.

Outputs:
- `z0Ptr`: A pointer to a 64-bit unsigned integer where the lower 64 bits of the product will be stored.
- `z1Ptr`: A pointer to a 64-bit unsigned integer where the upper 64 bits of the product will be stored.

Important Data Structures and Data Types:
- `bits64`: An unsigned 64-bit integer type used for the inputs and outputs of the multiplication.
- `bits32`: An unsigned 32-bit integer type used for intermediate calculations to split the 64-bit inputs into manageable parts.

Sub-Components:
- None
\end{prompt}
    \tcbsubtitle{kernel.h}
\begin{prompt}
typedef unsigned short int bits16;
typedef unsigned int bits32;
typedef unsigned long long int bits64;

void mul64To128(bits64 a, bits64 b, bits64 *z0Ptr, bits64 *z1Ptr);
\end{prompt}
    \tcbsubtitle{kernel.cpp}
\begin{prompt}
#include "mul64To128.h"

void mul64To128(bits64 a, bits64 b, bits64 *z0Ptr, bits64 *z1Ptr) {
    bits32 aHigh, aLow, bHigh, bLow;
    bits64 z0, zMiddleA, zMiddleB, z1;

    aLow = a;
    aHigh = a >> 32;
    bLow = b;
    bHigh = b >> 32;
    z1 = ((bits64)aLow) * bLow;
    zMiddleA = ((bits64)aLow) * bHigh;
    zMiddleB = ((bits64)aHigh) * bLow;
    z0 = ((bits64)aHigh) * bHigh;
    zMiddleA += zMiddleB;
    z0 += (((bits64)(zMiddleA < zMiddleB)) << 32) + (zMiddleA >> 32);
    zMiddleA <<= 32;
    z1 += zMiddleA;
    z0 += (z1 < zMiddleA);
    *z1Ptr = z1;
    *z0Ptr = z0;
}
\end{prompt}
    \tcbsubtitle{kernel\_tb.cpp}
\begin{prompt}
#include <cstdio>

#include "mul64To128.h"

int main() {
    int result = 0;

    struct {
        bits64 a;
        bits64 b;
        bits64 expected_high;
        bits64 expected_low;
    } test_cases[] = {
        {0x0000000100000000ULL,
         0x0000000100000000ULL,
         0x0000000000000001ULL,
         0x0000000000000000ULL},
        {0xFFFFFFFFFFFFFFFFULL,
         0xFFFFFFFFFFFFFFFFULL,
         0xFFFFFFFFFFFFFFFEULL,
         0x0000000000000001ULL},
        {0x123456789ABCDEF0ULL,
         0x0FEDCBA987654321ULL,
         0x0121FA00AD77D742ULL,
         0x2236D88FE5618CF0ULL},
        {0x0000000000000000ULL,
         0xFFFFFFFFFFFFFFFFULL,
         0x0000000000000000ULL,
         0x0000000000000000ULL},
        {0xFFFFFFFFFFFFFFFFULL,
         0x0000000000000001ULL,
         0x0000000000000000ULL,
         0xFFFFFFFFFFFFFFFFULL},
    };

    for (int i = 0; i < 5; ++i) {
        bits64 high, low;
        mul64To128(test_cases[i].a, test_cases[i].b, &high, &low);
        if (high != test_cases[i].expected_high ||
            low != test_cases[i].expected_low) {
            printf(
                "Test case %d FAILED: a = 0x%016llX, b = 0x%016llX, expected = "
                "(0x%016llX, 0x%016llX), got = (0x%016llX, 0x%016llX)\n",
                i,
                test_cases[i].a,
                test_cases[i].b,
                test_cases[i].expected_high,
                test_cases[i].expected_low,
                high,
                low);
            result = 1;
        }
    }

    if (result == 0)
        printf("All tests PASSED.\n");
    else
        printf("Some tests FAILED.\n");

    return result;
}
\end{prompt}
\end{tcolorbox}

\newpage
\subsection{Prompts for LLM-Aided Benchmark Construction}

As part of the benchmark construction, we have designed three meta-prompts to semi-automate the preparation of final benchmark designs from arbitrary HLS code sources as discussed in Section \ref{sec:benchmark_designs} and shown in Figure \ref{fig:bench-collection}. We show the prompts used for hierarchy extraction, natural language description generation, and testbench generation below.

\begin{tcolorbox}[title=Meta Prompt --- Hierarchy]
\begin{prompt}
You are a high-level synthesis expert.
Help me identify all the sub-components in this high-level synthesis hardware design.

Given the existing kernel code, header, and optional description, identify all the sub-components in the design.
A sub-component is a separate C++ function in the code that is not the top-level function.
This can also include sub-functions that are called by other sub-functions.
A sub-component should be identified by the function name.
If a sub-component is not listed as a separate C++ function, it should not be listed at all.
       
Output the list of sub-components in the design in a code block representing markdown.
    
The top level HLS kernel function is: `{top_name}`
       
Only output the list of sub-components in a code block representing markdown. Do not nest the list, just have a flat list of sub-components.
The output should be formatted exactly follows with no deviation:
       
```
- `subcomponent_1`
- `subcomponent_2`
- ...
```
       
Optional pre-existing simple kernel_description.md file:
    
{existing_description}
       
Input Kernel Code:
       
{kernel_code}
       
List of Sub-Components in requested markdown code block format (do not provide anything other than the list of sub-components in the requested format):
\end{prompt}
\end{tcolorbox}

\begin{tcolorbox}[title=Meta Prompt --- Description]
\begin{prompt}
You are a high-level synthesis expert.
Help me write a natural description of this high-level synthesis hardware design.

The description should cover the algorithm and functionality as well as the high level dataflow and architecture of the design.
Someone should be able to fully implement the design from the description provided.
Assume the reader knows little about the design and needs a detailed explanation.
Also include all details about any implementation quirks, edge cases, or design decisions that are important to this specific design.
This should include any latex equations if important or relevant to the design.
       
Include the list of top-level function inputs and outputs as well as a brief description of the functionality the kernel represents.
All arguments in the top-level function should be described in the inputs and outputs sections with details about the data type and layout.
Include a list of any important data structures and data types used in the design only if they are explicitly listed in the code.
Include a list of sub-components and a brief description of the functionality of each sub-component only if they are explicitly listed as separate C++ functions in the code.
If a sub-component is not listed as a separate C++ function, it should not be listed at all. 

Make sure descriptions about the high-level algorithm, inputs, outputs, data structures, and sub-components are detailed and thorough and include information about implementation, data type size layout, and architecture.
Each description can be multiple sentences long.

Sometimes a pre-existing kernel_description.md file is provided. If so use it as a starting guide but not the final output. Try to use all the information in the pre-existing description.

The top level kernel function is: `{top_name}`

Only output the description in a code block representing markdown.

The output should be formatted exactly follows with no deviation:
```
Kernel Description:
A high level natural language description of the design...
(be detailed and thorough, can be lengthy if needed, do not omit any details, can include latex equations if important or relevant to the design)

---

Top-Level Function: `name_of_top_level_function`

Complete Function Signature of the Top-Level Function:
`return_type name_of_top_level_function(input_1_type input_1, ...);`
        
Inputs:
- `input_1`: description of input_1...
- ....

Outputs:
- `output_1`: description of output_1...
- ....

Important Data Structures and Data Types:
- `data_structure_1`: description of data_structure_1... (description of the data structure, data type, size, layout, fields, use in the design, etc. are required)

Sub-Components:
- `subcomponent_1`:
    - Signature: `return_type subcomponent_1(input_1_type input_1, ...);`
    - Details: natural language description of subcomponent_1...
- ...
```

Optional pre-existing simple kernel_description.md file:

{existing_description}
       
Input Kernel Code:

{kernel_code}

Description in requested markdown code block format (do not provide anything other than the description in the requested format):
\end{prompt}
\end{tcolorbox}

\begin{tcolorbox}[title=Meta Prompt --- Testbench]
\begin{prompt}
You are a high-level synthesis expert.
Help me write a self contained single file testbench for this high-level synthesis hardware design.

This testbench should reside in a single C++ file and should be able to be compiled and run with the design to verify the correctness of the design.
The testbench should include a main function that calls the top-level function of the design with a set of test inputs and verifies the outputs.
The testbench should include a set of test inputs that are representative of the expected inputs to the design.
The testbench should include a set of expected outputs that are representative of the expected outputs of the design.
The inputs and outputs used should be enough to verify the functionality of the design.

The testbench should always return 0 if the design is correct and the testbench passes and should return a 1 if the design is incorrect and the testbench fails.

The current code may contain a testbench code and utility code for the testbench. If so, the goal is to simplify the testbench code and make it self-contained in a single file.
If the current code does not contain a testbench, the goal is to write a testbench from scratch that is self-contained in a single file.

If the current code contains a testbench, and the testbench uses precomputed data files in the testbench as golden inputs and outputs, the generated testbench must use them.

The top level HLS kernel function is: `{top_name}`

Only output the testbench code in a code block representing C++. No need to add the `cpp` identifier to the code block.

It should look like this:
    
```
testbench C++ source code
```
       
File List:
{file_list}
       
Optional pre-existing simple kernel_description.md file:

{existing_description}
       
Input Kernel Code:

{kernel_code}

Self-contained testbench code in requested C++ code block format (do not provide anything other than the testbench code in the requested format):
\end{prompt}
\end{tcolorbox}

\newpage

\subsection{Task Evaluation Prompts}

Below are the task prompts used in the various evaluations done in Section \ref{sec:evals}. These task prompts are combined with a larger prompt template, which we also include relevant snippets. For the full detail about how HLS code generation and HLS code editing prompts are constituted, please refer to our code.

\begin{tcolorbox}[title=Task Prompt --- HLS Kernel]
\begin{prompt}
## Task Description
Given a natural language description of an HLS design, a pre-written C++ design header, and a pre-written C++ testbench, generate the C++ implementation of the HLS design that aligns with the natural language description.

It should be functionally equivalent to the natural language description, be consistent with the provided header file, and pass the testbench. The design should also be synthesizable by the HLS tool.

Only generate the code for the design; do not modify the header file or the testbench. Make sure to import the header file as well.

Provide the complete design code in the single output; do not omit anything or leave placeholders.

Hierarchical design, sub-functions, template functions, structs, typedefs, and define statements are allowed but should be used only if appropriate.
\end{prompt}
\end{tcolorbox}

\begin{tcolorbox}[title=Task Prompt --- Loop Labels]
\begin{prompt}
### Editing Task - Loop Labels
Your task is to modify the given user's code to insert loop labels into the user's kernel (including in the kernel's top function, and any kernel subfunctions).
Only insert the loop labels; don't modify the actual loop code or insert any other pragmas.
Use the "labeled statement" C++ syntax as `label: statement` to label the loops.
If there are no loops in code, leave the code unchanged.
\end{prompt}
\end{tcolorbox}

\begin{tcolorbox}[title=Task Prompt --- Arbitrary and Fixed-Point Type Translation]
\begin{prompt}
## Editing Task - Arbitrary Precision and Fixed-Point Types
Your task is to modify the given user's code to convert the usage of int and uint types to arbitrary precision HLS types, `ap_int`, `ap_uint`, as well as convert float and double types to fixed-point HLS types, `ap_fixed`, provided by Vitis HLS.

- int and uint types should be converted to the appropriate arbitrary precision types, `ap_int` and `ap_uint`.
- float and double types should be converted to the appropriate fixed-point types, `ap_fixed`.

The integer types are defined as follows:
    - `ap_int<W>`: Signed integer type with `W` bits
    - `ap_uint<W>`: Unsigned integer type with `W` bits
In order to use ap_(u)int types, the user needs to include the "ap_int.h" header file.
The individual bits in the ap_(u)int types can be indexed using the [] operator.
    You can also set and clear bits at specific indexes in the ap_(u)int types using the set and clear methods:
        - void ap_(u)int::set (unsigned i)
        - void ap_(u)int::clear (unsigned i)

The fixed point type is defined as follows:
    `ap_fixed<W, I>`
where:
    - `W`: Word length in bits
    - `I`: The number of bits used to represent the integer value, that is, the number of integer bits to the left of the binary point, including the sign bit.
In order to use fixed point types, the user needs to include the "ap_fixed.h" header file.
The fixed point type can handle most C++ arithmetic operations (addition, subtraction, multiplication, division, etc.) and can be used in most C++ expressions.

If the user is also doing `cmath` operations on the original datatype numbers, these operations should be modified to use the HLS math library.
The HLS math library has the namespace `hls::*` and can be included with the following "hls_math.h" file. It supports most of the same math operators under the std::* namespace.

Typedefs for these new types are encouraged to make the code more readable.
Ideally, `typedef` statements should be placed in a header file so they can be reached by all source files.

The resulting code should maintain the same functionality as the original code but should convert all variants of int, uint, float, and double types to the appropriate arbitrary precision types.
\end{prompt}
\end{tcolorbox}

\begin{tcolorbox}[title=Task Prompt --- Dataflow Refactoring]
\begin{prompt}
## Editing Task - Dataflow Semantics
Your task is to modify the given user's code to use "dataflow" semantics in the HLS kernel using the `#pragma HLS DATAFLOW` pragma.
To use this pragma effectively, the code will need to be refactored into different subfunctions for computation tasks with intermediate producers and consumer variables.
Effectively, a dataflow function only contains calls to subfunctions as well as intermediate variables passed between the subfunctions.
The subfunctions must follow single-producer single-consumer rules, meaning that a variable / buffer can only be written to by one function and then only be read by one other function.
If data is needed for two subfunctions it needs to be duplicated.

The edited kernels must not have any of the following coding styles present in order to use dataflow semantics:
- Single-producer-consumer violations
- Feedback between tasks
- Conditional execution of tasks
- Loops with multiple exit conditions

In this case, if data needs to be buffered between tasks in a dataflow region, you may consider using fixed-sized arrays as buffers.

The resulting code should maintain the same functionality as the original code but should refactor the components into necessary dataflow subfunctions and use the DATAFLOW pragma in a manner which does not cause HLS synthesis to raise any dataflow violations.
\end{prompt}
\end{tcolorbox}

\begin{tcolorbox}[title=Task Prompt --- Loop Tiling]
\begin{prompt}
## Editing Task - Loop Tiling

Your task is to refactor the user's provided HLS kernel code by applying manual loop tiling source code transformations and inserting appropriate loop unrolling and array partitioning directives to optimize parallelism and memory access efficiency.

Specifically, you should:

- Manually tile loops into smaller blocks (tiles) with constant block sizes defined in the code to improve data locality and reduce memory bandwidth bottlenecks.

For a 2D example:

```
for(int i = 0; i < N; i++) {
    for(int j = 0; j < M; j++) {
        data[i][j] = ...;
        // loop body
    }
}
```

Should be transformed into:

```
const int N_TILE = 16;
const int M_TILE = 8;

#pragma HLS array_partition variable=data cyclic factor=N_TILE dim=1
#pragma HLS array_partition variable=data cyclic factor=M_TILE dim=2
<any more array partition pragmas needed for other variables>...

for(int i = 0; i < N; i += N_TILE) {
    for(int j = 0; j < M; j += M_TILE) {
        for(int ii = 0; ii < N_TILE; ii++) {
            #pragma HLS UNROLL
            for(int jj = 0; jj < M_TILE; jj++) {
                #pragma HLS UNROLL

                data[i + ii][j + jj] = ...;
                // loop body
            }
        }
    }
}
```

The same applies to 1D loops and n-D loops. Not all dimensions need to be tiled in every application.

- All loops must have fixed bounds and be perfect loops. Any other loop type is not allowed for tiling.
- The `#pragma HLS array_partition` pragma must be used if any arrays inside the loop are accessed using the tile indexes.
- Tiling with dependec on the outer loop is NOT allowed (ex. (ii = i; ii < i + N_TILE; ii++) is not allowed)
- Insert loop unrolling pragmas (`#pragma HLS UNROLL`) in the block loop of the loop tiling to minimize loop control overhead and maximize instruction-level parallelism.
- Be sure that the tiling factor is a constant value that is a factor of the loop trip count.
- If arrays are accessed using tile indexes, you must array partitioning pragmas (`#pragma HLS ARRAY_PARTITION <type> factor=... dim=...`) to allow parallel access to array elements inside unrolled loop tiles.
    - <type> can be `block` or `cyclic`, `factor` is the partitioning factor, and `dim` is the dimension of the array to partition (indexing starting at 1).
    - If multiple dims are partitioned, each dim of the array needs a separate pragma statement.
    - `factor` can be set to a const or define value in the code.
- Ensure the transformed loops maintain the same functionality and do not introduce loop-carried dependencies or initiation interval (II) violations.

If loop tiling and unrolling optimizations are not applicable to the provided kernel code, leave the code unchanged.
\end{prompt}
\end{tcolorbox}

\begin{tcolorbox}[title=General Prompt --- Output Format]
\begin{prompt}
## Output Format
The generated HLS edited output code should be provided in the following format:
```
<OUTPUT_CODE name="kernel_name.h">
    ...
</OUTPUT_CODE>
<OUTPUT_CODE name="kernel_name.cpp">
    ...
</OUTPUT_CODE>
<OUTPUT_CODE name="kernel_name_tb.cpp">
    ...
</OUTPUT_CODE>
```
Please use this XML format and do not use other formats like markdown code blocks or plain text.

You must output all three code blocks: the edited kernel code, the edited header file, and the edited testbench code.
If one of the files is not edited, you still need output the code block with the original code.
In the example above, `kernel_name` should be replaced with the original name of the kernel.
Make sure the testbench filename ends with `_tb.cpp`.

Only output the generated HLS code in the XML format and nothing else.
\end{prompt}
\end{tcolorbox}

\begin{tcolorbox}[title=General Prompt --- Preamble]
\begin{prompt}
## Overview
You are a helpful export hardware engineer and software developer who will assist the user with hardware design tasks for high-level synthesis.
The task will center around high-level synthesis (HLS) code written in C++ for a hardware design. The HLS design is written to target the latest Vitis HLS tool from Xilinx, which maps C++ code to a Verilog implementation for FPGAs.
\end{prompt}
\end{tcolorbox}

\end{document}